# Effects of electrostatic screening on the conformation of single DNA molecules confined in a nanochannel

**Ce Zhang**

*Biophysics and Complex Fluids Group, Department of Physics, National University of Singapore, 2 Science Drive 3, Singapore 117542*

**Fang Zhang and Jeroen A. van Kan**

*Centre for Ion Beam Applications, Department of Physics, National University of Singapore, 2 Science Drive 3, Singapore 117542*

**Johan R. C. van der Maarel**[a]

*Biophysics and Complex Fluids Group, Department of Physics, National University of Singapore, 2 Science Drive 3, Singapore 117542*

May 7, 2008

---

[a] Author for correspondence Tel: 65 65164396; Fax: 65 67776126; E-mail: phyjrcvd@nus.edu.sg

Single T4-DNA molecules were confined in rectangular-shaped channels with a depth of 300 nm and a width in the range 150-300 nm casted in a poly(dimethylsiloxane) nanofluidic chip. The extensions of the DNA molecules were measured with fluorescence microscopy as a function of the ionic strength and composition of the buffer as well as the DNA intercalation level by the YOYO-1 dye. The data were interpreted with scaling theory for a wormlike polymer in good solvent, including the effects of confinement, charge, and self-avoidance. It was found that the elongation of the DNA molecules with decreasing ionic strength can be interpreted in terms of an increase of the persistence length. Self-avoidance effects on the extension are moderate, due to the small correlation length imposed by the channel cross-sectional diameter. Intercalation of the dye results in an increase of the DNA contour length and a partial neutralization of the DNA charge, but besides effects of electrostatic origin it has no significant effect on the bare bending rigidity. In the presence of divalent cations, the DNA molecules were observed to contract, but they do not collapse into a condensed structure. It is proposed that this contraction results from a divalent counterion mediated attractive force between the segments of the DNA molecule.

## I. INTRODUCTION

In recent years, micro- and nanofluidic devices have been developed and utilized for the analysis of the properties of long biopolymers including DNA. Different architectures have been designed, including entropic trap arrays, micro and nanopillar arrays, nanopores and nanochannels.[1,2,3,4] The nanofluidic devices are complementary to other single molecule manipulation techniques such as those based on optical tweezers.[5,6,7] In the case of confinement in a long and straight nanochannel, a long biomolecule elongates because of the restriction in configurational degrees of freedom imposed by the channel walls. This effect is similar to the situation in a single molecule stretching experiment, in which the biomolecule is subjected to a tensional force. The role of the strength of the stretching force is the same as the value of the cross-sectional diameter of the nanochannel, in the sense that both a stronger force and a decrease in diameter result in a more extended conformation. Besides the similarities of these single-molecule manipulation techniques, there are also some important differences. For instance, in the confinement experiment there is no need for chemical modification to attach the biomolecule to molecular pincers. Furthermore, the stretching data are usually interpreted with the wormlike Gaussian chain model.[8] The Gaussian chain model does not include the effects of self-avoidance or, in other words, excluded volume effects. In the analysis of the confinement experiment, the latter effects are included in a straightforward way, just as in the case of the description of the swelling of a single polymer coil in a good solvent.

The statistics of a DNA molecule confined in a nanochannel depends on the ratio of the cross-sectional channel diameter and the DNA persistence length as well as self-avoidance.[9,10] In a narrow channel, with a width less than the persistence length, the molecule will undulate inside the channel and will only bend when it bounces off the wall.[11] In the present contribution, the channel diameters exceed the persistence length, but they are smaller than the radius of gyration of the unconstrained, free DNA coil. In the latter situation, the elongated DNA molecule inside the channel remains coiled at all length scales.[9] The advantage of such configuration is that the data can be interpreted using well established polymer theory, including the effects of the local bending rigidity (persistence length) and interaction of spatially close segments which are separated over a long distance along the contour (excluded volume). The scaling relations for the extension of DNA molecules confined in nanochannels of various diameters have been verified by Reisner *et al*.[12] The effects of electrostatic screening on the conformation of DNA molecules inside nanoslits and nanochannels of various dimensions also have been reported before.[13,14] In general, the DNA molecule has been observed to stretch out with decreased screening of Coulomb interaction. The variation in extension with the ionic strength of the supporting medium was interpreted in terms of a variation in the persistence length. Contradictive results were reported however for the relative importance of excluded volume effects. Accordingly, we thought it of interest to reinvestigate the effects of electrostatic screening and, in particular, to gauge the relative importance of self-avoidance with respect to the bending rigidity for the understanding of the conformational response to changes in ionic environment.

Most, if not all of the previously reported experiments on DNA were done using nanofluidic devices made out of fused silica. We have used a lithography process with proton beam writing to fabricate a nanopatterned stamp.[15,16] The stamp was subsequently replicated in poly(dimethyl siloxane) (PDMS) polymer, followed by curing and sealing with a glass slide.[17,18] In order to visualize the DNA molecules with fluorescence microscopy it is necessary to stain them with a dye. We have used YOYO-1, which bis-intercalates between three adjacent base-pairs with a concomitant increase in DNA contour length. The maximum level of intercalation is four base-pairs per dye molecule. Furthermore, YOYO-1 carries four positive charges, so that it also partially neutralizes the DNA phosphate charge (for maximum loading the DNA charge is thus reduced by a factor of two). We have systematically investigated the effect of the YOYO-1 intercalation level on the DNA conformation. In particular, we have explored the minimum level of staining in order to obtain the structural DNA properties as close to the native state as possible and, yet, to achieve a sufficiently intense fluorescence signal for accurate imaging. At higher staining levels, we will show that the effect of intercalation can be accounted for by the partially compensating effects of the increase in contour length and the decrease in persistence length due to the reduced net DNA charge.

Most of our experiments were done using conventional buffers, which primarily contain monovalent ions. In order to investigate a possible dependence of the DNA conformation on the ionic composition of the buffer, we have done experiments using both a high (Tris-HCl) and a low (Tris-boric acid) ionic strength solvent medium. Furthermore, we have investigated the effects of divalent magnesium, calcium, and the polyamine putrescine on the extension of the DNA molecules inside the nanochannels. The changes in DNA conformation in response to changes in ionic environment were evaluated using screened electrostatics, based on an effective

DNA charge density following from the solution of the non-linear Poisson–Boltzmann equation for a rodlike polyelectrolyte immersed in an excess of mono- or divalent salt. As we will see shortly, for the (primarily) monovalent buffer systems this procedure gave quite satisfactorily results. In the case of the divalent ions significant discrepancies were observed, which cannot be reconciled with the theory for a wormlike polymer in a good solvent. We tentatively explain these discrepancies in terms of a divalent counterion mediated attraction between DNA segments, which is not predicted by the conventional mean-field theory.

## II. THEORY

### A. Confinement in a nanochannel

The conformation of a DNA molecule confined in a channel depends on, among others, the channel diameter $D$ and the DNA persistence length $L_p$.[9,10,19] In a narrow channel with $D < L_p$ the molecule will be aligned and takes a highly extended conformation. In the present contribution, the channel diameters are in the range 150 to 300 nm, whereas the persistence lengths are between, say 50 and 150 nm. Furthermore, the size of the free DNA coil, as characterized by its radius of gyration $R_g$, is larger than the channel diameter so that $L_p < D < R_g$ ($R_g = 1.4$ µm for T4-DNA in 10 mM salt). In this situation, the DNA molecule coils inside the channel, but it will be anisotropic with a certain elongation in the longitudinal direction. As we will see shortly, the elongation depends on the channel diameter and screening conditions in the supporting buffer medium. For now, we will ignore specific interaction between the DNA molecule and the wall of the channel. All derived properties of the confinement are thus related to the restriction in configurational degrees of freedom and screening of excluded volume interactions.

In de Gennes' scaling approach, the DNA molecule inside the channel is considered as a linear sequence of sub-coils of uniform size.[9] These sub-coils are commonly referred to as blobs.

Within the blobs the DNA molecule is thought to be unaffected by the presence of the channel walls. Furthermore, excluded volume effects are screened beyond the blob size, because segments pertaining to different blobs can not form contact pairs. Successive blobs have a second virial coefficient of repulsive interaction of the order of their volume and, hence, they are packed into a linear array (for a discussion of the free energy of confinement see below). The size of the blobs is characterized by their radius of gyration $R_b$, which should be on the order of the cross-sectional diameter of the channel (do not confuse $R_b$ with the radius of gyration of the free DNA coil $R_g$). In the scaling treatment, the exact relationship between $R_b$ and $D$ is unknown. We simply assume a linear dependence $D = C R_b$ with $C$ being an adjustable parameter on the order of unity. The parameter $C$ is universal, in the sense that its value should not depend on the chain statistics inside the blobs. It might include however certain wall-induced effects such as depletion at the wall interface.[20] If the contour length stored in a single blob is given by $L_b$, the total number of blobs in the aligned sequence of blobs is $L/L_b$ with $L$ being the total contour length of the DNA molecule. The average relative extension in the longitudinal direction of the channel then takes the form $R_\parallel/L = D/L_b = C R_b/L_b$. Each blob fluctuates with a second moment of its distribution of mass $\langle \delta R_b^2 \rangle = R_b^2$. For a string of $L/L_b$ independently fluctuating blobs, each with a second moment in size $C^2 R_b^2$, the second moment of the extension is given by $\langle \delta R_\parallel^2 \rangle = C^2 R_b^2 L/L_b$.

The relationship between the blob size $R_b$ and the contour length $L_b$ depends on the chain statistics. Inside the blobs the DNA chain is swollen due to excluded volume interactions. Self-avoidance can be taken into account by defining a swelling factor $\alpha_s$, which relates the radius of gyration of the swollen blob to the one pertaining to the hypothetical unperturbed Gaussian blob $R_b = \alpha_s R_b^0$. The swelling factor $\alpha_s$ is related to the excluded volume expressed in terms of the parameter[21]

$$z = \frac{3}{16}\left(\frac{3}{\pi}\right)^{1/2} \left(\frac{L_b}{L_p}\right)^{1/2} D_{eff}/L_p .$$ (1)

The parameter $z$ is proportional to the scaled effective diameter of the DNA molecule $D_{eff}/L_p$ and the square root of the number of persistence length segments inside the blob $\left(L_b/L_p\right)^{1/2}$. For DNA confined in nanochannels, the values of $z$ are typically less than 0.45. For such small values of $z$, the swelling factor $\alpha_s$ is given by the series expansion up to and including second order in $z$:[22]

$$\alpha_s^2 = 1 \ + \ 1.276 \ z \ - \ 2.082 \ z^2 . \tag{2}$$

For a long Gaussian chain, the unperturbed radius of gyration of the coil is given by $\left(R_b^0\right)^2 = L_b L_p/3$. However, due to the relatively small number of persistence length segments, inside the blob the limiting long chain expression does not suffice. It is our contention that for quantitative calculations it is essential to use the full Benoit-Doty equation for the unperturbed radius of gyration of the relatively short section of the DNA molecule within the blob[23]

$$\left(R_b^0\right)^2 = \frac{L_p}{3L_b^2}\left[L_b^3 - 3L_b^2 L_p + 6L_b L_p^2 - 6L_p^3\left(1 - \exp\left(-\frac{L_b}{L_p}\right)\right)\right]. \tag{3}$$

For a given channel diameter $D$, persistence length $L_p$, and effective diameter $D_{eff}$, the contour length stored in the blob $L_b$ can be obtained by numerically solving $D = \alpha_s C R_b^0$. The average of the relative extension then follows from

$$R_{\parallel}/L = D/L_b = \alpha_s C \, R_b^0/L_b \tag{4}$$

with the values for $\alpha_s$ and $R_b^0$ given by Eq. (2) and (3), respectively. The width of the distribution in relative extension following from a large number of measurements of individual molecules takes the form

$$\left\langle \delta R_{\parallel}^2 \right\rangle^{1/2}\big/L = D\big/\left(LL_b\right)^{1/2} = \alpha_s C R_b^0\big/\left(LL_b\right)^{1/2} . \tag{5}$$

It is interesting to compare the extension of a DNA molecule confined inside a nanochannel with the one of a molecule subjected to a tensional force $f$ in a tweezers set-up. In the single molecule stretching experiment, the blob size is determined by the balance of the elastic stretching energy and the thermal energy $fR_b = kT$. With $D = CR_b$ we obtain $f = C \, kT/D$. With $C$ of the order of unity, it follows that confinement inside a channel with a cross-sectional diameter of about 200 nm corresponds with a stretching force of about 0.02 pN.

For a polymer chain in a relatively wide channel, the free energy of confinement can be derived using a scaling argument.[9,10] The free energy should be an extensive property and, hence, it should be proportional to the length of the polymer. Furthermore, it should have the dimension of thermal energy and it should depend on the ratio of the two relevant length scales of the system, $i.e.$, the channel diameter $D$ and the radius of gyration $R_g$ of the free, unconstrained polymer coil under the same buffer conditions. A free DNA coil of a large molecular weight is characterized by large values of the excluded volume parameter $z$, for which the swelling parameter is given by the modified Flory result $\lim_{z \to \infty} \alpha_s^5 = 1.276 \, z$ [Eq. (2) is only valid for small values of $z$ and is used for the statistics inside the blobs]. Accordingly, the radius of gyration of the free DNA coil is given by the Flory expression $R_g \simeq \left(L_p D_{eff}\right)^{1/5} L^{3/5}$ with a proportionality factor around 0.75. The scaling relation for the free energy of confinement then takes the form

$$F_{conf} \simeq kT \left(R_g/D\right)^{5/3} \qquad (6)$$

where the exponent $5/3$ was chosen for a linear dependence on the contour length $L$. The repulsive energy per blob follows from $DF_{conf}/R_{\parallel}$ and should be constant and on the order of the thermal energy $kT$.

## B. DNA as a polyelectrolyte

DNA is a polyelectrolyte and both the persistence length and the effective diameter depend on the electrostatic screening conditions of the supporting buffer medium.[10] In the Debye-Hückel approximation, the effective diameter of a rodlike persistence length segment takes the form

$$D_{eff} = D_0 + \kappa^{-1}\left(\ln w' + \gamma - 1/2 + \ln 2\right) \qquad (7)$$

with $w' = w \exp(-\kappa D_0)$ and $w = 2\pi \nu_{eff}^2 l_B \kappa^{-1}$.[10,21] Here, $D_0$ is the bare diameter (2 nm for DNA in the *B*-form), $\gamma$ denotes Euler's constant, and $\nu_{eff}$ is the effective number of charges per unit contour length of the DNA molecule. The electrostatic screening length $\kappa^{-1}$ is given by $\kappa^2 = 8\pi l_B \mu$ with $\mu$ being the ionic strength of the solvent medium and the Bjerrum length $l_B = e^2/4\pi\varepsilon kT$. Electrostatic interaction also modifies the bending rigidity of the DNA molecule, *i.e.* its persistence length. In the Odijk-Skolnick-Fixman theory, the total persistence length

$$L_p = L_p^0 + L_p^e \qquad (8)$$

is given by the sum of the bare and the electrostatic part.[24,25] For DNA in the *B*-form, the value of the bare persistence length is around 50 nm.[5,6,7] The electrostatic part also has been derived in the Debye-Hückel approximation by considering the electrostatic energy cost of bending a wormlike chain

$$L_p^e = \nu_{eff}^2 l_B \big/ \left(4\kappa^2\right). \qquad (9)$$

The electrostatic persistence length depends on the screening conditions of the supporting medium through the electrostatic screening length $\kappa^{-1}$ as well as the effective number of charges per unit contour length of the DNA molecule $\nu_{eff}$.

The key parameter in the calculation of the effective diameter and the electrostatic contribution to the persistence length is $\nu_{eff}$. The latter parameter was obtained from the numerical solution of the non-linear Poisson-Boltzmann equation for a cylindrical polyelectrolyte in excess monovalent and/or divalent salt using a fourth order Runge-Kutta method.[10,26] Note that the value of $\nu_{eff}$ depends on the screening conditions of the supporting medium due to the high non-linearity in the inner double layer region. In our fluorescence imaging experiments we have stained the DNA molecules with the YOYO-1 dye. Each YOYO-1 molecule carries four positive charges, so that the net DNA charge is substantially reduced for higher degrees of intercalation (our highest employed intercalation level was four base-pairs per YOYO-1 molecule). Furthermore, YOYO-1 increases the contour length of the DNA molecule by 0.4 nm per dye molecule.[27,28,29] Both the charge neutralization effect and the increase of the contour length by the staining dye were included in our calculations of the effective diameter and electrostatic persistence length.

Fig. 1 displays the calculated $D_{eff}$ and $L_p$ as a function of the ionic strength of the supporting medium. In accordance with our experimental conditions, these calculations were done for an intercalation level of 23 base-pairs per YOYO-1 molecule and for a buffer composed of monovalent ions only. In the case of a buffer with added divalent salt, we have obtained similar results by solving the non-linear Poisson-Boltzmann equation for the relevant mixture of mono- and divalent ions. With decreasing ionic strength and hence decreased screening of electrostatic interaction both the persistence length and the effective diameter increases. Eventually, for very low ionic strength less than, say, 0.3 mM the values of these structural parameters become on the order of the width of our narrowest channel (150 nm). For a fixed ionic strength of the supporting buffer ($\mu = 0.61$ mM), the effect of the level of intercalation is displayed in Fig. 2. With increased staining levels (less base-pairs per YOYO-1 molecule) both $D_{eff}$ and $L_p$ decreases due to concomitant charge neutralization and the increase of the DNA contour length (decrease in linear DNA charge density). Note that in the theoretical calculations the value of the bare persistence length was kept fixed at 44 nm in accordance with the experimental observations to be discussed below. All variations in the structural parameters come from variation in the buffer screening conditions and the DNA charge density through intercalation of the YOYO-1 dye.

## III. EXPERIMENTAL PROCEDURES

### A. Fabrication of the nanofluidic chip

The nanofluidic device was made by replication in PDMS of a patterned master stamp. The stamp was fabricated in HSQ resist (Dow Corning, Midland, MI) using a lithography process with proton beam writing.[15,16] A scanning electron microscopy image of the HSQ stamp bearing the positive nanostructures is displayed in Fig. 3a. The image shows two channel wall pairs with 150 and 200 nm width, respectively (a third channel structure with 300 nm width is outside the field of view). The 300 nm height of the positive channel structures on the stamp was measured with atomic force microscopy (Dimension 3000, Veeco, Woodbury, NY). The stamp was replicated in PDMS followed by curing with a curing agent (Sylgard, Dow Corning).[17,18] An atomic force microscopy image of a section of the negative PDMS replica showing 200 nm wide channels is displayed in Fig. 3b. Finally, the PDMS replica was sealed with a glass slide after both substrates have been plasma oxidized (Harrick, Ossining, NY). The widths of the channels in the PDMS replica were measured with atomic force microscopy and the values agreed with those obtained from the scanning electron microscopy images of the HSQ master stamp. The nanochannels were rectangular-shaped and have a depth of $300 \pm 5$ nm and a width of $150 \pm 5$, $200 \pm 5$, and $300 \pm 5$ nm, respectively. More details about the chip design and fabrication will be published elsewhere.

## B. DNA sample preparation

T4 GT7 DNA (T4-DNA, 165.65 kbp) and λ-phage DNA (λ-DNA, 48.502 kbp) were purchased from Nippon Gene, Tokyo and New England Biolabs, Ipswich, MA, respectively, and used without further purification. YOYO-1 fluorescence staining dye was purchased from Invitrogen, Carlsbad, CA. Water was deionized and purified by a Millipore system and had a conductivity less than $1 \times 10^{-6}$ $\Omega^{-1}$cm$^{-1}$. Samples were prepared by dialyzing solutions of T4-DNA against the relevant buffer solutions in micro-dialyzers. The buffer systems were Tris-borate/EDTA (TBE), Tris/HCl (T), or TBE with added $MgCl_2$, $CaCl_2$ or putrescine-$Cl_2$ (TBE/Mg, TBE/Ca, or TBE/Pu, respectively). After dialysis, the T4-DNA was stained with YOYO-1 and diluted with the relevant buffer to a final DNA concentration of 0.008 g/L. The dye intercalation ratio was 4.0, 5.7, 12, 15, 23, or 46 base-pairs per YOYO-1 molecule. In TBE, the final buffer concentrations were 1/100, 1/75, 1/50, 1/30, 1/15, 1/3, and 1×TBE (1×TBE is 90 mM Tris, 90 mM boric acid, and 2 mM EDTA, pH 8.5). For the relatively high ionic strength T buffer system, the concentrations were 1/10, 1, and 10×T (1×T is 10 mM Tris adjusted with HCl to pH 8.5). The TBE/Mg, TBE/Ca, and TBE/Pu buffer systems were composed of 1/100×TBE with 0.09, 0.56, and 3.33 mM added $MgCl_2$, $CaCl_2$, or putrescine-$Cl_2$. We also prepared a matching sample pair of T4-DNA and λ-DNA with an intercalation ratio of 4.0 base-pairs per YOYO-1 molecule immersed in 5×TBE buffer. The ionic strengths and buffer compositions were calculated with the Davies equation for estimating the activity coefficients of the ions. The utilized pKa's are 10.17, 6.11, 2.68, 2.00 for EDTA, 13.80, 12.74, and 9.24 for boric acid, and 8.08 for Tris.

## C. Fluorescence imaging

The YOYO-1-stained DNA molecules dispersed in the relevant buffer were loaded into two reservoirs connected by the nanochannels. The DNA molecules were subsequently driven into the channels by an electric field. For this purpose, two platinum electrodes were immersed in the reservoirs and connected to an electrophoresis power supply with a voltage in the range 10-30 V (Bio-Rad, Hercules, CA). Once the DNA molecules were localized inside the channels, the electric field was switched off and the molecules were allowed to relax to their equilibrium state for at least 60 seconds. The fluorescence of the stained DNA molecules were visualized with an Olympus IX71 inverted fluorescence microscope equipped with a 100 W mercury lamp, a UV filter set and a 100 times oil immersion objective. The exposure time of the YOYO-1 fluorescence was controlled by a UV light shutter. Images were collected with a charge coupled device (CCD) camera (Olympus DP 70) and the extension of the elongated DNA molecules inside the channels was measured with the public domain software ImageJ (http://rsb.info.nih.gov/ij/). Reported relative extensions, *i.e.* extensions divided by the contour length of the DNA molecule, represent an average of at least 200 individual measurements. A montage of fluorescence images of T4-DNA in 300×300 nm$^2$ channels under various TBE buffer conditions is shown in Fig. 4.

## IV. RESULTS AND DISCUSSION

### A. Effect of contour length

An implication of the scaling approach is that the extension should be proportional to the contour length of the DNA molecule [see Eq. (4)]. A linear scaling of the extension with the contour length has been reported before.[30] In order to check our experimental procedures and to reconfirm the linear scaling we have measured the extensions of T4 and λ-DNA with an equal intercalation ratio of 4.0 base-pairs per YOYO-1 molecule and immersed in 5×TBE buffer. In the 200×300 nm$^2$ channel, the extensions have the values $8.1 \pm 2.4$ and $2.5 \pm 0.5$ μm for T4 and λ-DNA, respectively (in the 300×300 nm$^2$ channel slightly smaller values were obtained). Note that the standard deviations refer to the width of the distribution following from the measurement of at least 200 different DNA molecules for each data point. Based on the linearity of the extension in the contour length, the ratio of the extensions in the nanochannels should match the ratio of the DNA contour lengths. The contour length of T4–DNA exceeds the one of λ–DNA by a factor 3.41, which is close to the experimental ratio $3.2 \pm 1.2$ of the corresponding extensions. Accordingly, the extension is indeed proportional to the contour length, which supports the use of the linearly packed blob model for a polymer chain in a good solvent.[9]

### B. Effects of ionic strength

The DNA molecules were stained with the YOYO-1 dye in order to image them with fluorescence microscopy. We did a series of experiments with a relatively low level of intercalation of 23 base-pairs per YOYO-1 molecule. This level of intercalation results in a sufficiently intense fluorescence for imaging (see Fig. 4). Yet, the distortion of the secondary

DNA structure is minimal; the contour length has increased from 56.6 to 59.5 micrometer and the DNA charge is reduced by a factor 42/46 only.[28,29] Furthermore, we have employed two different solvents: the Tris-borate/EDTA (TBE) and Tris/HCl (T) buffer systems. TBE buffer contains some multivalent ions (divalent boric acid and di-, tri-, and tetravalent EDTA), but at pH 8.5 their concentrations are vanishingly small. The T buffer system contains monovalent ions only. We measured the extensions of individual T4-DNA molecules confined in the channels with three different cross-sections: 300×300, 200×300, and 150×300 nm$^2$. The relative extensions are set out in Fig. 5 as a function of the ionic strength of the buffer medium. For both buffer systems, the experimental results match, irrespective the channel cross-section and ionic strength. Notice that the molecules significantly elongate below an ionic strength of around 5 mM, but they remain coiled with maximum extensions of about 35% of their contour length. Our results are qualitatively similar to those reported before for $\lambda$-phage DNA in fused silica channels and slits of comparable dimensions.[12,13]

The theoretical extension Eq. (4) was fitted to the data in Fig. 5 by optimizing the values for the constant $C$ and the bare persistence length $L_p^0$. The effective DNA diameters and electrostatic persistence lengths were calculated as described in the theoretical section. For the relevant buffer and YOYO-1 staining conditions the effective diameter and the total persistence lengths are displayed in Fig. 1. In order to take into account the effect of the electrostatic interaction of the DNA molecule with the negatively charged channel wall, we have used a reduced channel diameter by subtracting two times the electrostatic screening length from the bare cross-sectional diameter. Furthermore, for asymmetric channel cross-sections we have taken the square root of the depth times the width as the effective channel cross-sectional diameter. All data are well reproduced with the optimized values for the constant $C = 1.45$ and bare persistence length $L_p^0 = 44$ nm. The fitted value of $L_p^0$ agrees with the commonly accepted value for DNA in the $B$-form of around 50 nm.[5] The value of the constant $C$ is indeed of the order unity and there is no significant dependence on the channel cross-section and solvent ionic strength. Furthermore, the experimental widths of the distribution in relative extension agrees with the theoretically predicted bandwidth Eq. (5).

In the theoretical procedure, the elongation of the DNA molecule with decreased screening of Coulomb interaction is interpreted in terms of changes in local bending rigidity of the DNA duplex (persistence length) and self-avoidance effects of interacting rod-like DNA segments inside the blobs (excluded volume effects). The strong increase in persistence length with decreasing solvent ionic strength is illustrated in Fig. 1. As a result of the increase in persistence length, the number of persistence length segments per blob $L_b/L_p$ decreases with decreasing ionic strength (see Fig. 6). At very low ionic strength, less than 0.3 mM, the number of persistence length segments per blob becomes very small of the order of three. Due to the relatively small number of persistence length segments per blob and, in particular, in the case of low ionic strengths, it is essential to use the full Benoit-Doty Eq. (3) rather than the long Gaussian chain result for the unperturbed radius of gyration of the chain section within the blob. Note that excluded volume effects are accounted for through the swelling factor $\alpha_s$ (see below). The blob model becomes inapplicable when the value of the persistence length surpasses the channel cross-sectional diameter, but this situation does not occur under the present experimental conditions. The repulsive energy per blob $DF_{conf}/R_\parallel$ has been calculated with the help of Eq. (6) and is also set out in Fig. 6. In the 300×300 nm² channel the repulsive energy is almost constant in the range of ionic strength below 5 mM where the DNA molecule significantly elongates. The small residual ionic strength dependence of $DF_{conf}/R_\parallel$ for higher ionic stengths and/or smaller channel cross-sectional diameters is related to deficiencies of the model such as an imperfect estimation of the swelling factor or wall-induced effects on the excluded volume.

It is interesting to gauge the relative importance of the swelling of the blobs by excluded volume effects in the interpretation of the ionic strength dependence of the extension data. Excluded volume effects were accounted for by using the series expansion of the swelling factor Eq. (2), which is only valid for small values of the parameter $z$. With decreasing ionic strength, the effective diameter increases, but so does the persistence length (see Fig. 1). Furthermore, the number of persistence length segments per blob decreases with decreased screening of Coulomb interaction. As a result of these partially compensating effects, the value of the excluded volume parameter $z$ remains small and exhibits a maximum at an intermediate ionic strength around 2 mM. The values of $z$ and the resulting swelling factor $\alpha_s$ are displayed in Fig. 7. The swelling factor shows a very small variation about the mean value 1.087 within a bandwidth of $\pm 0.008$. Since according to Eq. (4) the extension is proportional to $\alpha_s C$, the data could almost equally well be fitted without the consideration of excluded volume effects, *i.e.*, by setting $\alpha_s = 1$ and with a 9 % larger value of the parameter $C$. Note that this does not imply that the DNA molecule behaves as a random Gaussian chain and/or that self-avoidance effects are unimportant. Irrespective the channel cross-sectional diameter, the repulsive energy per blob remains on the order of *kT* and the blobs are packed in a linear array (see Fig. 6).

The ionic strength dependence of our extension data can hence be explained by a variation of the persistence length alone. This finding is at odds with the earlier reported conclusion by Reisner *et al*., who explicitly stated that it is necessary to include self-avoidance effects in order to apprehend the conformational response of single and confined DNA molecules to changes in the ionic environment.[14] The latter authors also interpreted their results with the scaling model of de Gennes, but they used the Flory radius for the blob size which is exclusively valid for *long* chain segments. We have used the relevant combination of the unperturbed radius of gyration and swelling factor for shorter chains, in accordance with the relatively small correlation lengths imposed by the channel cross-sectional diameters. This difference in the description of the chain statistics within the blob explains the discrepancy between our and the previously reported conclusion about the significance of self-avoidance effects in the interpretation of the ionic strength dependence of the extension of single DNA molecules confined in a nanochannel.

## C. Effects of intercalation

The above discussed experiments were done employing a relatively low level of intercalation of 23 base-pairs per YOYO-1 molecule. This low level of intercalation ensures a minimal distortion of the DNA secondary structure and net charge. In order to gauge the effect of intercalation on the conformation, we have measured the relative extension of T4-DNA with the saturating intercalation level of four base-pairs per YOYO-1 molecule as a function of the ionic strength of the supporting TBE medium. It should be noted that at the saturating intercalation level the contour length of the T4-DNA molecule has increased from 56.6 to 73.2 micrometer and the net DNA charge is reduced by a factor of two (the spine axis projected distance between charges of the labeled T4-DNA is thus increased from 0.171 to 0.442 nm).[28,29] The results obtained in channels of a cross-section of 200×300 nm$^2$ are displayed in Fig. 8. For the other channel cross-sections similar results are obtained (data not shown). With the saturating level of intercalation, the relative extensions are somewhat smaller than the values obtained with an intercalation level of 23 base-pairs per YOYO-1 molecule.

The theoretical extension Eq. (4) was fitted to the data in Fig. 8 by optimizing the values for the constant $C$ and the bare persistence length $L_p^0$. The effective diameters and persistence lengths were also obtained by solving the non-linear Poisson Boltzmann equation, but with the relevant linear charge density pertaining to the highly intercalated DNA molecule. The optimized values for the fit parameters are $C = 1.43$ and $L_p^0 = 44$ nm, which are in excellent agreement with the results presented above obtained for T4-DNA with an intercalation level of 23 base-pairs per YOYO-1 molecule. As a surprising result, we find that the bare persistence length is insensitive to the level of intercalation. This insensitivity is at odds with the reported DNA persistence length of as short as 12 nm obtained by single molecule optical tweezers experiments on λ-phage DNA stained with the YOYO-1 dye with a much higher intercalation level of 0.73 base-pair to YOYO-1 ratio.[31] An explanation for this discrepancy is that at such low base-pair to dye ratio not all dye molecules can be accommodated in the regular way of one YOYO-1 molecule bis-intercalated between three adjacent base-pairs. A fraction of the YOYO-1 molecules might be intercalated in an irregular way (such as multiple intercalated dye molecules per stretch of four base-pairs) or bound sideways to the duplex. These irregular forms of intercalating and binding of the 'surplus' of YOYO-1 molecules are expected to have a profound effect on the persistence length, since they disrupt the base-pair stacking interaction to which DNA largely owes its bending rigidity.

The effect of intercalation on the DNA conformation at a constant buffer concentration of 1/50×TBE is demonstrated in Fig. 9. Here, the theoretical extension Eq. (4) was fitted to the data by only optimizing $C = 1.39$ and by using the effective diameters and persistence lengths as displayed in Fig. 2. The optimized values of the constant $C$ obtained from the various data sets are all very close with an average value $C = 1.42 \pm 0.03$. This shows that $C$ is indeed a universal constant and does not depend on the screening conditions or net DNA charge. The variation of the extension with the degree of intercalation is reasonably well reproduced by the theoretical prediction using a constant value of the bare persistence length of $L_p^0 = 44$ nm. With decreasing level of intercalation (higher base-pair to YOYO-1 ratio), the contour length decreases by 0.4 nm per dye molecule.[28,29] Concurrently, the linear charge density increases, which results in an increase of the total persistence length and effective diameter (see Fig. 2). As a result of these two partially compensating effects, the *relative* extension of the DNA molecule confined in the channel first shows a modest increase after it levels off at an ionic strength dependent, but constant value for intercalation levels exceeding 10 base-pairs per YOYO-1 molecule.

### D. Effects of divalent salt

It is well known that cations of valence three or greater can induce a condensed structure, *i.e.* a highly packaged state of single or multiple DNA molecules with an ordered morphology.[32,33,34] Divalent cations can also induce condensation provided the dielectric constant of the medium is reduced by the addition of alcohol.[35,36] Here, we have investigated the conformation of the T4-DNA molecules confined in the nanochannels in the presence of divalent magnesium, calcium, or putrescine. In order to drive the DNA molecules inside the channels, we found that the solvent needs to be buffered. Accordingly, we have measured the relative extension in $1/100 \times$ TBE with various amounts of added $MgCl_2$, $CaCl_2$, or putrescine-$Cl_2$. The intercalation ratio was 23 base-pairs per YOYO-1 molecule, so the distortion of the secondary DNA structure was minimal. The experimental results obtained with a channel cross-section of $300 \times 300$ nm$^2$ are displayed in Fig. 10. With increasing ionic strength generated by the addition of the divalent salt, the relative extension reduces. The molecules are however not condensed into a compact structure. Furthermore, the results for the various divalent cations are quite similar, but in the presence of the linear polyamine putrescine the chain takes a somewhat more extended conformation.

The theoretically predicted extension based on the solution of the non-linear Poisson-Boltzmann equation for the relevant mixture of mono- and divalent ions is shown in Fig. 10 (calculated with $C = 1.45$ and $L_p^0 = 44$ nm). For comparison we also have included the optimized theoretical prediction pertaining to the extension of T4-DNA under TBE buffer conditions without added divalent salt. The experimental values fall clearly below the mark set by the theoretical prediction, in particular for the higher ionic strengths. We verified that in the presence of divalent cations Eq. (4) is unable to reproduce the extension data with a swelling factor larger than unity. With increasing divalent salt concentration the DNA molecule contracts with a more compact statistics than the swollen chain. Note that for a polymer chain in a theta or poor solvent the scaling model is no longer applicable. Nevertheless, the results show that divalent cations give rise to an attractive force among the segments of a long DNA molecule in a single coil, but apparently the force is not strong enough to induce condensation. An attractive force mediated by divalent counterions between very *short* DNA fragments (8-25 base pairs) was recently observed in a small angle X-ray scattering (SAXS) study of the second virial coefficient.[37] In the presence of monovalent salt the second and third virial coefficients of persistence length DNA fragments agree with a repulsive interaction and screened electrostatics.[38] Full atomic scale Molecular Dynamics simulations have revealed attractive ion bridges, *i.e.*, multivalent ions which are simultaneously and temporarily bound to opposing DNA duplexes with a lifetime on the order of a few nanoseconds.[39] Furthermore, a small angle neutron scattering (SANS) study of the synthetic poly(styrenesulfonic acid) (PSS) has shown that the chain contracts below the Gaussian chain limit in the presence of calcium counterions.[40] The similarity in behavior of PSS and DNA suggests that the multivalent ion mediated attraction is a general phenomenon pertaining to highly charged linear polyelectrolytes.

## V. CONCLUSIONS

Single DNA molecules can be confined and extended in long and straight nanochannels casted in a PDMS nanofluidic chip. The cross-sectional diameters of the channels were in the range 150-300 nm, which are larger than the persistence length, but smaller than the radius of gyration of the free, unconstrained T4-DNA coil. Inside the channel the DNA molecule hence remains coiled at all length scales, but it is extended in the longitudinal direction due to the restriction imposed by the channel walls. With the help of fluorescence microscopy, the DNA molecules were visualized and their extensions were measured. It was checked that the extension is proportional to the contour length of the DNA molecule. The relative extensions were interpreted with a modification of de Gennes' scaling theory for a polymer chain in a good solvent including the effects of charge. Because of the fact that the correlation length is relatively small, it is essential to use the full Benoit-Doty equation for the unperturbed radius of gyration of the relatively short chain segment within the blob rather than the commonly used expression for the long Gaussian chain. Self-avoidance effects were included by using the perturbation series expansion of the coil swelling factor in terms of the excluded volume parameter. Due the relatively small number of statistical segments within the blob, the values of the excluded volume parameter are small and the chain statistics is only mildly perturbed with a value of the swelling factor around 1.09. Furthermore, the latter value was found to be rather robust against a variation in ionic strength. The variation of the extension of the confined DNA molecule with a change in ionic strength of the supporting buffer medium can be described by a variation of the DNA persistence length induced by screening of the Coulomb interaction. Note that successive blobs remain packed in a linear array with a repulsive energy per blob on the order of the thermal energy $kT$.

In order to visualize the DNA molecules with the help of fluorescence microscopy, it is necessary to stain them with an intercalating dye. Intercalation of the YOYO-1 dye molecule results in an increase of the contour length as well as a partial compensation of the DNA charge. We observed a moderate decrease in the relative extension with increasing level of intercalation. The extensions could be interpreted using the polymer scaling theory, including the effects of the intercalating dye on the secondary DNA structure as well as the partial DNA charge neutralization. All extension data agree with a single value of the bare persistence length $L_p^0 = 44$ nm, irrespective the intercalation level and the screening conditions of the supporting medium. Although YOYO-1 increases the contour length of the DNA molecule, our experiments show that it has no significant effect on the bending rigidity (at least in the intercalation level range of 46 to four base-pairs per dye molecule). Furthermore, we found that the radius of gyration of the blob is related to the channel diameter through a constant. This constant $C = D/R_b$ was found to be universal with an average value $C = 1.42 \pm 0.03$, in the sense that it is independent on the DNA charge (degree of intercalation), screening conditions, and channel cross-sectional diameter. The effect of the electrostatic interaction of the DNA molecule with the plasma oxidized negatively charged channel wall was accounted for by using a reduced channel diameter through the subtraction of two times the electrostatic screening length from the bare cross-sectional diameter.

In the presence of divalent counterions, the relative extension also reduces with increased screening of Coulomb interaction. Here, the experimental values of the relative extensions fall however clearly below the mark set by the theoretical prediction including the effects of divalent ions on the screened electrostatics. The extension results imply that the DNA molecule inside the channel contracts with a more compact statistics than the swollen chain. This change in conformation is proposed to result from divalent counterion mediated attractive bridging interactions between the DNA segments. These interactions are counterion specific, because in the case of the linear putrescine the DNA molecule was observed to take a slightly more extended conformation than in the presence of the small and spherical calcium and magnesium counterions. The divalent ions do however not induce condensation and, in this respect, their behavior is similar to that in the bulk phase.

**Acknowledgement**. The support of grant R-144-000-177-112 from the Singapore Ministry of Education is acknowledged. We thank Jean-Louis Sikorav for discussions.

**Legends to the figures**

Fig. 1    Persistence length $L_p$ (solid curve) and effective diameter $D_{eff}$ (dashed curve) versus the ionic strength $\mu$ of the buffer medium calculated with the effective number of charges per unit DNA length from the numerical solution to the non-linear Poisson-Boltzmann equation for DNA in the *B*-form and 23 base-pairs per YOYO-1 molecule (bare diameter $D_0 = 2$ nm and bare persistence length $L_p^0 = 44$ nm).

Fig. 2    Persistence length $L_p$ (solid curve) and effective diameter $D_{eff}$ (dashed curve) versus the number of base-pairs per YOYO-1 molecule calculated with the effective number of charges per unit DNA length from the numerical solution to the non-linear Poisson-Boltzmann equation for DNA in the *B*-form in 1/50×TBE buffer (ionic strength $\mu = 0.61$ mM).

Fig. 3    (a): Scanning electron microscopy image of the HSQ master showing a positive standing channel structure with a width of 150 (I) and 200 (II) and a uniform height of 300 nm (the height was determined by atomic force microscopy). Feature III is a 500 nm wide 'feeder' channel. (b): Atomic force microscopy image of the PDMS replica showing a set of parallel channels with a width and depth of 200 and 300 nm, respectively.

Fig. 4    Montage of fluorescence images of T4-DNA molecules confined in 300×300 nm² channels. The buffer concentration decreases from left to right according to 1, 1/15, 1/30, 1/50, 1/75, and 1/100×TBE. The molecules were stained with an intercalation level of 23 base-pairs per YOYO-1 molecule.

Fig. 5  Relative extension $R_\parallel/L$ of T4-DNA in 150×300 (a), 200×300 (b), and 300×300 (c) nm$^2$ channels versus the ionic strength $\mu$ of the solvent medium. The open symbols refer to TBE buffer conditions, whereas the closed symbols pertain to the T buffer system. The error bars denote the width of the distribution following from the measurement of at least 200 different DNA molecules for each data point. The solid curves represent a fit of the theoretically predicted extensions to the data with the persistence lengths and effective diameters displayed in Fig. 1 and optimized scaling constant $C = 1.45$ and bare persistence length $L_p^0 = 44$ nm. The dashed curves represent the fluctuation bandwidth. The molecules were stained with an intercalation level of 23 base-pairs per YOYO-1 molecule.

Fig. 6  (a) Number of persistence length segments per blob $L_b/L_p$ in 150×300 (dashed-dotted curve), 200×300 (dashed curve), and 300×300 (solid curve) nm$^2$ channels versus the ionic strength $\mu$ of the solvent pertaining to the fit of the theoretical extension to the data in Fig. 5. (b) As in panel (a), but for the repulsive energy per blob $DF_{conf}/R_\parallel$ in units $kT$.

Fig. 7  (a) Excluded volume parameter $z$ in 150×300 (dashed-dotted curve), 200×300 (dashed curve), and 300×300 (solid curve) nm$^2$ channels versus the ionic strength $\mu$ of the solvent pertaining to the fit of the theoretical extension to the data in Fig. 5. (b) As in panel (a), but for the swelling factor $\alpha_s$.

Fig. 8  Relative extension $R_\parallel/L$ of T4-DNA stained with the saturating intercalation ratio of 4 base-pairs per YOYO-1 molecule in the 200×300 nm$^2$ channel versus the ionic strength $\mu$ of the TBE medium. The solid curve represents a fit of the theoretically predicted extension calculated with the relevant linear DNA charge density and optimized scaling

constant $C = 1.43$ and bare persistence length $L_p^0 = 44$ nm. The dashed curve is included for comparison and represents the corresponding theoretical result for T4-DNA with 23 base-pairs per YOYO-1 molecule (see Fig. 5).

Fig. 9   Relative extension $R_\parallel / L$ of T4-DNA versus the base-pair to YOYO-1 ratio in 1/50×TBE buffer. The solid curve represents a fit of the theoretically predicted extension to the data with the persistence lengths and effective diameters shown in Fig. 2 (optimized with a fixed value $C = 1.39$).

Fig. 10   Relative extension $R_\parallel / L$ of T4-DNA versus the ionic strength of the TBE/Mg (□), TBE/Ca (○), and TBE/Pu (△) buffers in 300×300 nm$^2$ channels. The intercalation ratio is 23 base-pairs per YOYO-1 molecule. The solid curve represents the theoretically predicted extension based on the solution of the non-linear Poisson-Boltzmann equation including the effect of the divalent salt. The dashed curve is included for comparison and represents the corresponding result for the relative extension in a monovalent salt buffer (see Fig. 5).

**Figure 1**

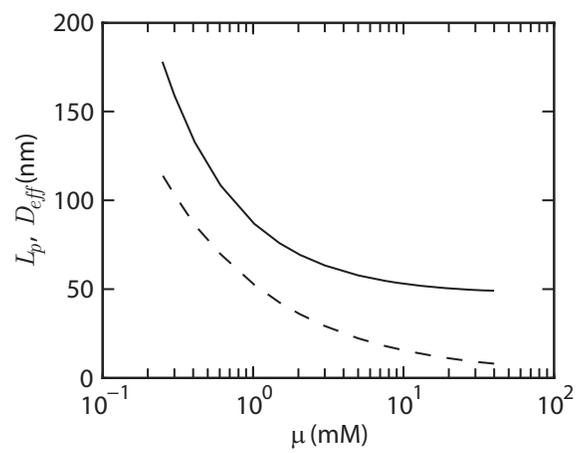

**Figure 2**

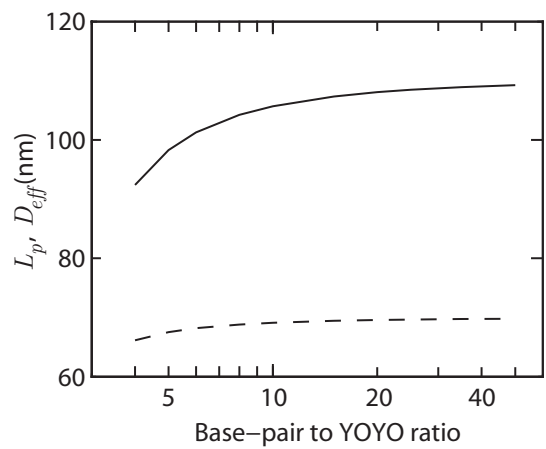

**Figure 3**

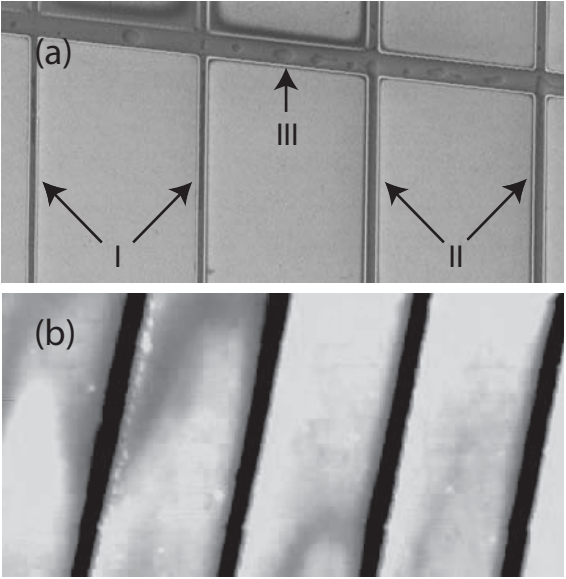

**Figure 4**

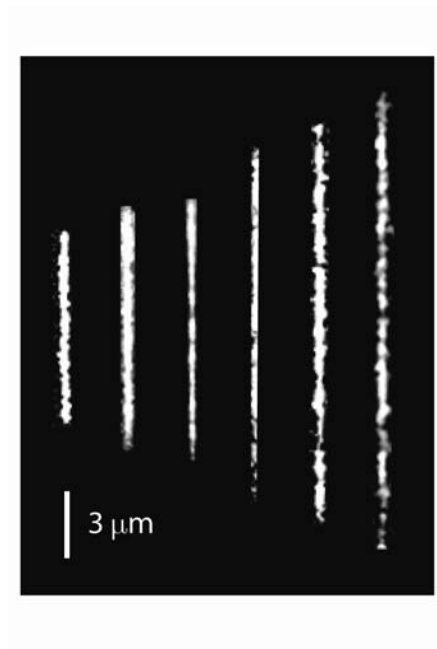

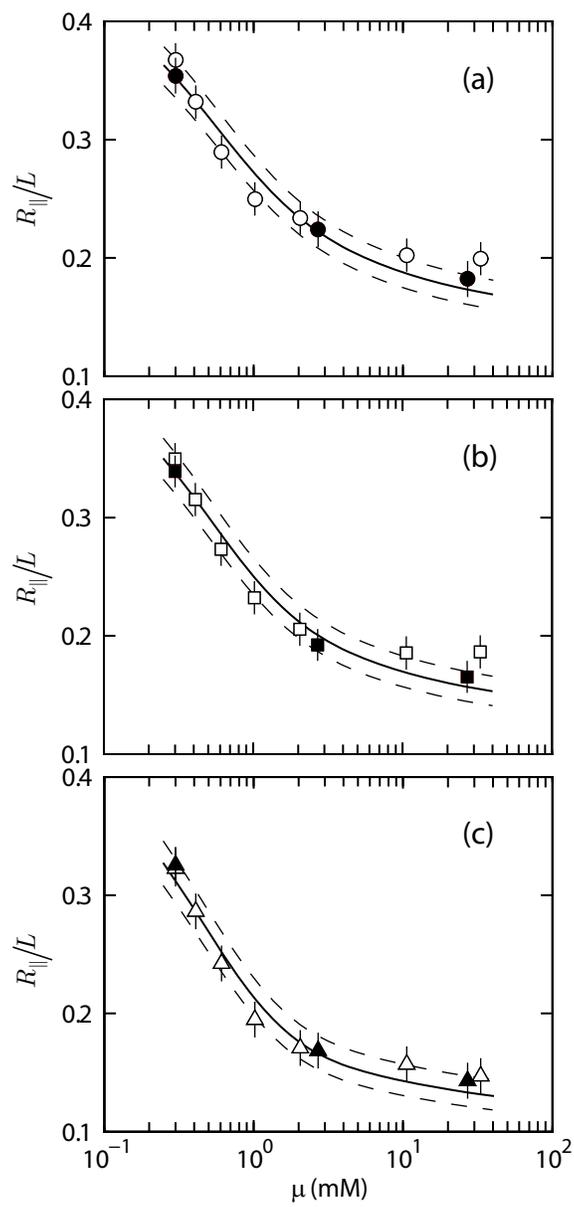

**Figure 6**

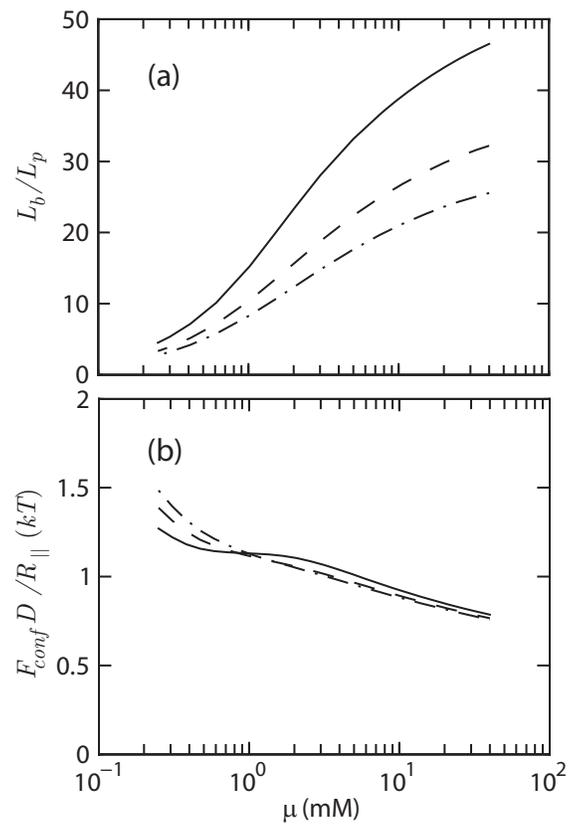

**Figure 7**

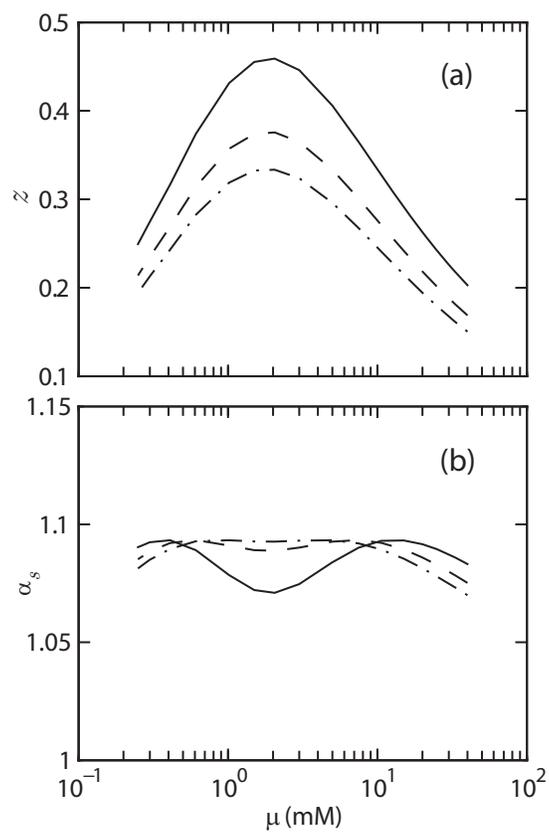

**Figure 8**

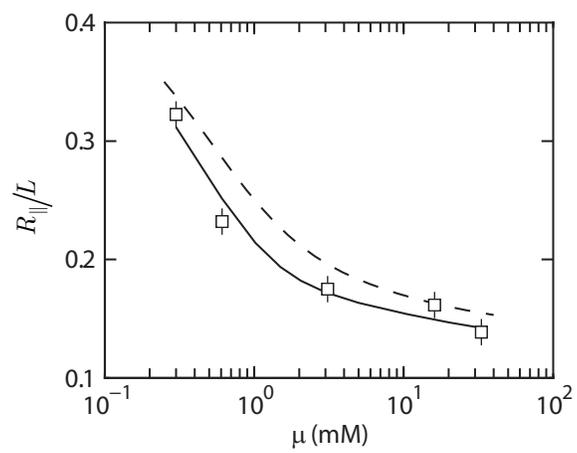

**Figure 9**

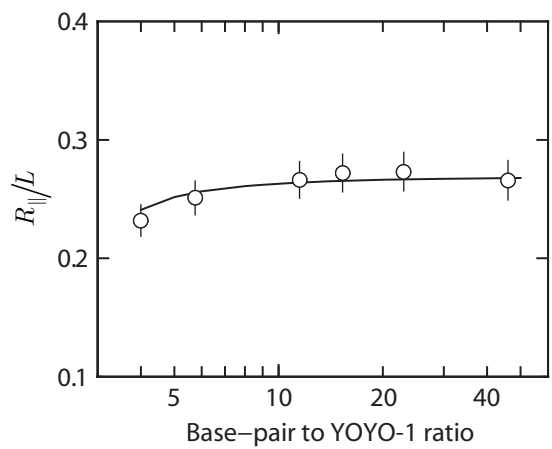

**Figure 10**

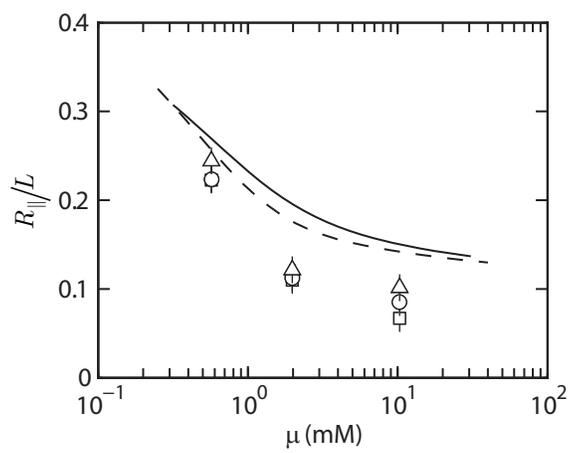